\documentclass[pre,aps,superscriptaddress,showpacs,floatfix,preprint]{revtex4-1}
\usepackage{amssymb,mathrsfs,amsfonts}
\usepackage{amsmath}
\usepackage{graphicx}
\usepackage{epstopdf}
\usepackage{nicefrac}
\usepackage{listings}
\usepackage{upgreek}
\usepackage{physics}
\usepackage{multirow}
\usepackage{color}
\usepackage{textcomp}
\usepackage{bibnames}
\usepackage{appendix}
\usepackage{epstopdf}
\usepackage{nicefrac}
\usepackage{listings}
\usepackage{upgreek}
\usepackage{physics}
\usepackage{multirow}
\usepackage{color}
\usepackage{natbib}
\usepackage{graphicx}
\usepackage{tabls}
\usepackage{afterpage}
\usepackage{amsthm}
\usepackage{lastpage}
\usepackage{empheq, etoolbox}

\usepackage{epstopdf}
\usepackage{changes} %
\colorlet{Changes@Color}{red}  %change the color of package of changes

\newcommand{\A}{{\cal A}}
\newcommand{\weff}{{\omega_{\mathrm{eff}}}}
\newcommand{\B}{{\cal B}}

\newcommand{\RN}[1]{%
  \textup{\uppercase\expandafter{\romannumeral#1}}%
}

\newcommand{\hI}{\hat{I}_{s}}
\newcommand{\hE}{\hat{E}_{s}}
\newcommand{\hsigma}{\hat{\sigma}^{s}}
\newcommand{\halpha}{\hat{\alpha}_{s}}

\newcommand{\hkappa}{\hat{\varkappa}_{s,0}}

\newcommand{\hweff}{{\hat{\omega}_{\mathrm{eff}}}{}_{,s}}
\newcommand{\hDN}{\hat{D}_{s,N}}

\begin{document}
\title{Asymptotic $P_N$ Approximation in Radiative Transfer Problems} 
\author{Re'em Harel}
\affiliation{Department of Physics, Bar-Ilan University, 5290002, Ramat-Gan, Israel}
\author{Stanislav Burov}
\affiliation{Department of Physics, Bar-Ilan University, 5290002, Ramat-Gan, Israel}
\author{Shay I. Heizler}
\email{Shay.Heizler@mail.huji.ac.il}
\affiliation{Racah Institute of Physics, The Hebrew University, 9190401 Jerusalem, Israel}

\begin{abstract}
We study the validity of the time-dependent asymptotic $P_N$ approximation in radiative transfer of photons. The time-dependent asymptotic $P_N$ is an approximation which uses the standard $P_N$ equations with a closure that is based on the asymptotic solution of the exact Boltzmann equation for a homogeneous problem, in space and time. The asymptotic $P_N$ approximation for radiative transfer requires careful treatment regarding the closure equation. Specifically, the mean number of particles that are emitted per collision ($\weff$) can be larger than one due to inner or outer radiation sources and the coefficients of the closure must be extended for these cases. Our approximation is tested against a well-known radiative transfer benchmark. It yields excellent results, with almost correct particle velocity that controls the radiative heat-wave fronts. 
\end{abstract}

\maketitle

\section{Introduction} 
\label{intro}
The Boltzmann (transport) equation is an integro-differential equation which describes the local density of particles traveling and interacting inside a media~\cite{Duderstadt,CaseZweifel1967}. In high energy density physics phenomena, it appears in radiative transfer and describes the absorption and emission of photons within physical materials~\cite{pomraning2005equations}. Since the radiation opacity and the heat capacity of materials depend strongly on the material's temperature, it generates nonlinear waves, i.e. Marshak waves. The physics that is governed by the radiative transfer equation (RTE) is used for the study of many physical phenomena, for example, modeling theoretical astrophysics (supernova)~\cite{castor2004,pomraning2005equations}, laboratory plasma and inertial confinement fusion (ICF)~\cite{lindl2004,rosen1996}.

There are several ways to solve the transport equation, among them are the spherical harmonics method (the $P_N$ approximation)~\cite{pomraning2005equations}, the discrete ordinates method (the $S_N$ method)~\cite{sn} and Monte-Carlo simulations~\cite{imc}. Of course, the Monte-Carlo method is a statistical model while the first two solvers are deterministic transport equations and tend to the exact solution only for infinite set of coupled equations ($N\to\infty$). However, in some cases solving exactly the transport equation is impractical and resource-consuming. Thus, different approximations were introduced to reduce the complexity of the problem. The most well-known approximation, the {\em classic diffusion approximation}, can be developed by assuming that the angular dependence is isotropic or close to isotropic, and it is a simplification of the $P_N$ using $N=1$, (assuming steady-state behavior of the particle current)~\cite{pomraning2005equations}. Thus, the diffusion approximation tends to the exact solution in the limit of local thermodynamic equilibrium (LTE), close to the steady-state. As a result, one pitfall of the diffusion approximation is that it fails to describe the particles' density in highly anisotropic scenarios, specifically in optically thin media~\cite{su1997analytical,Olson1999}. Flux-Limiters (in a diffusion-like equation) or Variable Eddington Factors (in full $P_1$ equations) may be introduced, however these approximations are nonlinear making the solution hard to obtain for multi-dimensional cases, and yield unstable solutions in specific situations~\cite{Su2001}. 

In the $P_N$ approximation the specific intensity is expanded in a full set of spherical harmonics thus, transforming the transport equation to a set of coupled equations, with a cutoff method or closure~\cite{Duderstadt,CaseZweifel1967,pomraning2005equations}. For example, the classical $P_N$ approximation is obtained by introduction of a closure that sets to zero all the coefficients of the expansion with an index larger than $N$~\cite{Duderstadt,CaseZweifel1967,pomraning2005equations}. Alternative $P_N$ closure was introduced by Pomraning~\cite{pomraning1964generalized, pomraning1965asymptotically} for time-independent problems. It was derived as a general-$N$ expansion of the asymptotic diffusion approximation~\cite{case1953introduction}, based on the spatial asymptotic time-independent solution for the exact Boltzmann equation. The physical justification for the asymptotic distribution of particles to fit well in various finite systems with sufficient number of mean free paths (MFP) comes from the physics of neutron transport. The asymptotic diffusion yields the correct critical radii, even for systems with less than two mean MFPs, while the classical $P_N$ approximation converges slowly compared to the exact solution~\cite{case1953introduction, BellGlasstone}.

For time-dependent problems, a related approximation was proposed by Heizler~\cite{heizler2010asymptotic}, namely the {\em asymptotic $P_1$ approximation} that is based on an asymptotic behavior in both space and time of the transport equation. This approximation is, in some sense, a time-dependent equivalent to the asymptotic diffusion approximation. This approximation has a $P_1$ form that overcomes the parabolic nature of the asymptotic diffusion. It yields the correct time-independent eigenvalue of the exact transport equation, and produces velocity that is very close to the correct one. In homogeneous problems, the asymptotic $P_1$ yields better results than the asymptotic diffusion, and much better than the classic $P_1$ approximation. This approximation was derived and tested also for local LTE radiative transfer~\cite{heizler2012}, and shares similar asymptotic behavior as $P_2$ approximation~\cite{heizler2012sp}. Further derivation for heterogeneous problems was introduced by the discontinuous asymptotic $P_1$ approximation~\cite{cohen2018discontinuous, cohen2019discontinuous}, which is important especially in radiative transfer problems, when the cross-sections are strong functions of the material's temperature.

Recently a new time-dependent asymptotic $P_N$ approximation was introduced~\cite{harel2020time}, which is a generalization of the work with the asymptotic $P_1$, and is based on the time-independent closure proposed by Pomraning while using time-asymptotic analysis. This approximation was tested in homogeneous media problems and yields very good results, including the {\em{tails}} of the distribution, even for low-order $N$. In this paper we generalize the testing to heterogeneous problems, specifically for radiative transfer and utilizing the Su-Olson benchmark~\cite{su1997analytical}. To avoid redundancy with~\cite{harel2020time}, we present only short summary of the derivation of the time-dependent asymptotic $P_N$ approximation, with  attention to the special modifications that were done to include radiative transfer problems.

First, in Section~\ref{section_pn} we present the $P_N$ for  both the classic and the Pomraning closure methods. Afterwards we derive our approximation named the time-dependent asymptotic $P_N$ approximation in its radiative transfer form. Then, in Sec.~\ref{test_benchmark} we compare our method to the classical $P_N$ approximation and the Su-Olson benchmark~\cite{su1997analytical}.

\section{$P_N$ Equations} \label{section_pn}

In this section, we present a derivation of the time-dependent asymptotic $P_N$ approximation in its radiative transfer form. First, we present the simple form of the $P_N$ equations with the classic $P_N$ closure, and the time-independent asymptotic closure proposed by Pomraning~\cite{pomraning1964generalized}. Next, a new time-dependent $P_N$ approximation, that is based on the a time-asymptotic analogy of Pomraning's closure, is described. This derivation is based on~\cite{harel2020time}, while taking special care of the new modifications that have to be made for radiative transfer.

\subsection{The Classic $P_N$ Approximation}

In one-dimensional slab geometry, the gray radiative transfer Boltzmann equation attains the form:
\begin{equation} \label{transport_Eq}
    \frac{1}{c}\pdv{I(x, \mu, t)}{t} + \mu \pdv{I(x, \mu, t)}{x} + \sigma_t I (x, \mu, t) = \sigma_a B(T_m(x,t))+ \frac{1}{2}\int_{-1}^{1} \sigma_s(\mu_0)I(x, \mu', t) d\mu' + S(x, t)
\end{equation}
Where $I(x, \mu, t)$ is the specific intensity of radiation in position $x$ at time $t$ in $\mu$ direction, i.e., the cosine of the angle with respect to the $x$-axis ($\mu=\Omega \cdot \hat{x}$). $c$ is the speed of light, $B(T_m(x,t))$ is the local black-body radiation emitted from the material, where $T_m(x,t)$ is the material temperature. $\sigma_t=\sigma_s+\sigma_a$ is the total cross section which is composed of $\sigma_s(\mu_0)$ (where $\mu_0\equiv\Omega \cdot \Omega'$), the scattering cross section (such as Thomson or Compton scattering) and $\sigma_a$, the absorbing cross section (the opacity). $S(x, t)$ is the external source term. The equation for the radiation is coupled to the complementary equation for the material temperature, $T_m(x,t)$: 
\begin{equation}\label{materialequation}
    \frac{C_v(T_m(x,t))}{c} \pdv{T_m(x,t)}{t} = \sigma_a \left(\frac{1}{c}\int_{-1}^{1} I(x, \mu, t)d\mu - a T_m^4(x,t)\right)
\end{equation}
where $C_v(T_m(x,t))$ is the material's heat capacity and $a$ is the radiation constant and defined by $a\equiv 4\sigma_{\mathrm{sb}}/c$, and $\sigma_{\mathrm{sb}}$ is the Stefan-Boltzmann constant ($B(x,t) = aT_m^4$). 

To derive the various $P_N$ approximations, we expand the specific intensity ($I(x,\mu,t)$) in a full set of spherical harmonics~\cite{Duderstadt,CaseZweifel1967,pomraning2005equations}, or in the one-dimensional case to a full set of Legendre polynomials:
\begin{subequations}
    \begin{equation}\label{polynomseries}
        I(x, \mu, t) = \sum_{n=0}^{\infty}\frac{2n+1}{2} I_n(x, t) P_n(\mu),
    \end{equation}
    \begin{equation}
    I_n(x,t) = \int_{-1}^{1} P_n(\mu) I(x, \mu, t)d\mu
\end{equation}
\end{subequations}
where $P_n(\mu)$ are the Legendre polynomials and the angular moments are given by $I_n(x,t)$. The first two moments are $I_0(x, t)\equiv cE(x, t)$ - the energy density and $I_1(x, t)\equiv F(x, t)$ - the radiation flux. Plugging Eq.~\ref{polynomseries} back into the transport equation (Eq.~\ref{transport_Eq}) and using the orthogonal property of the Legendre polynomials yields infinite coupled set of equations where each equation is given by:
\begin{equation}\label{pnequations}
\frac{1}{c} \pdv{I_n(x, t)}{t} + (\sigma_t-\sigma_s^{(n)}) I_n(x, t) +\left(\frac{n+1}{2n+1}\right)  \pdv{I_{n+1}(x, t)}{x} + \left(\frac{n}{2n+1}\right) \pdv{I_{n-1}(x, t)}{x} = S^{(n)}(x, t)
\end{equation}
$\sigma_s^{(n)}$ is the $n$'th moment of the scattering cross-section and $S^{(n)}(x, t)$ is the $n$'th moment of the external source term. Both of these terms can be derived in the same way that $I_n(x,t)$ is derived, i.e by expanding them into Legendre series.

In order to reduce the infinite sum to a finite sum some sort of closure needs to be introduced. As an example, for the classical $P_N$ approximation the closure is simply setting all the coefficients with index larger than $N$ to zero:
\begin{equation}
    \pdv{I_{n+1}(x, t)}{x}=0 \qquad\qquad n\geqslant N.
\end{equation}
This results in $N-1$ \textbf{exact} equations, that can be expressed with Eqs.~\ref{pnequations} and the last $N$'th \textbf{approximated} equation is defined by: 
\begin{equation}\label{pn_classic_closure}
    \frac{1}{c}\pdv{I_N(x,t)}{t} + (\sigma_t - \sigma_s^{(N)})I_N(x,t) + \left(\frac{N}{2N+1}\right)\pdv{I_{N-1}(x, t)}{x} = S^{(N)}(x,t).
\end{equation}
Eqs.~\ref{pn_classic_closure} and~\ref{pnequations} yield a closed set of $N+1$ equations for $N+1$ unknown moments. The case of $N=1$ tends to the gray diffusion approximation, when neglecting the derivative of $F(x, t)$ (the radiation flux) with respect to time. The derivation, from here on, will use the notation of isotropic scattering term. For the case of anisotropic scattering, one have to replace $\sigma_t$ with $\sigma_t - \sigma_s^{(n)}$.

\subsection{Pomraning's Time-independent Asymptotic $P_N$}\label{pomramingpna}
Pomraning offered a new closure that is based on the time-independent asymptotic particles distribution due to the exact Boltzmann equation in an infinite, source-free homogeneous medium, which can be solved analytically~\cite{pomraning1964generalized}. As a matter of fact, this is a generalization of the asymptotic diffusion approximation~\cite{case1953introduction,CaseZweifel1967} (using $N=1$), for a general $N$. Pomraning originally derived this approximation in its neutron form however, we will derive it in its radiative transfer form. The last approximated equation takes the following form:
\begin{equation}\label{pomraningtruncation}
        \sigma_tI_N(x) + \left(\frac{N}{2N+1}\right)\pdv{I_{N-1}(x)}{x} + \left(\frac{N + 1}{2N + 1}\right) \pdv{(\alpha_N I_{N-1}(x))}{x} =0
\end{equation}
where $\alpha_N$ is the term that attempts to retain the symmetry of the transport equation~\cite{pomraning2005equations}, and is defined as:
\begin{equation}\label{alphapomraning0}
 \alpha_N = \frac{\int_{-1}^{1} I_a(x,\mu) P_{N+1}(\mu) d\mu}{\int_{-1}^{1} I_a(x,\mu) P_{N-1}(\mu) d\mu}
\end{equation}
$I_a$ is the spatial asymptotic form of the specific intensity. As one can notice, by setting $\alpha_N$ to zero, we obtain the classical $P_N$ approximation. In the general case, $I_a$ can be derived in multiple ways, each suitable for some angular distribution. Specifically, the asymptotic case can be derived from the exact solution of the time-independent, source-free Boltzmann equation for infinite homogeneous medium~\cite{case1953introduction,CaseZweifel1967}:
\begin{equation}\label{boltzmman_indep}
    \mu \pdv{ I(x,\mu)}{x} = \sigma_t \left(\frac{\weff }{2}cE(x) - I(x,\mu)\right)
\end{equation}
Where $\weff$ is the mean number of particles that are emitted per collision or source terms and is defined by (generalized to time-dependent case)~\cite{pomraning2005equations,cohen2018discontinuous}:
\begin{equation}
    \weff(x,t) = \frac{\sigma_s E(x,t) + \sigma_a B(x,t) + S(x,t)/c}{\sigma_t E(x,t)}.
\label{omega}
\end{equation}
$\weff$ is the radiative transfer equivalence of the albedo $c$ that was used in~\cite{harel2020time} (not to be confused with the speed of light, in this paper), that is common mostly for neutronic modeling. Unlike in  our previous work~\cite{harel2020time}, in the case of strong sources $\weff$ may be much larger than unity (as an in multiplying media), which forces additional derivations for this work. Also notice that in the case of  LTE, $\weff\equiv 1$,  the asymptotic distribution tends to the diffusion limit. 

By a separation of variables~\cite{case1953introduction,CaseZweifel1967}, one can solve the transport equation~\ref{boltzmman_indep} and obtain a closed transcendental equation for the eigenvalues of the positional asymptotic solution - $\varkappa_0$, as a function of $\weff$: 
\begin{equation}
    \frac{2}{\weff}=\frac{1}{\varkappa_0}\ell n \left(\frac{1+\varkappa_0}{1-\varkappa_0}
    \right)
\label{kappa}
\end{equation}
and the asymptotic eigenfunctions - $I_a$ are (after forcing normalization): 
\begin{equation}\label{asymptotc_eq}
I(x,\mu)\approx I_a(x,\mu)=\frac{\weff}{2}A_0\frac{e^{\varkappa_0\sigma_tx}}{1+\mu\varkappa_0}+
\frac{\weff}{2}B_0\frac{e^{-\varkappa_0\sigma_tx}}{1-\mu\varkappa_0}
\end{equation}

Plugging the asymptotic solution (Eq.~\ref{asymptotc_eq}) into the closure term (Eq.~\ref{alphapomraning0}), yields:
\begin{equation}\label{alphapomraning}
\alpha_N = \frac{\int_{-1}^{1} \left(\frac{1}{1+\mu\varkappa_0}\right) P_{N+1}(\mu) d\mu}{\int_{-1}^{1} \left(\frac{1}{1+\mu\varkappa_0}\right) P_{N-1}(\mu) d\mu}
\end{equation}
Plugging Eq.~\ref{alphapomraning} back into Eq.~\ref{pomraningtruncation} yields Pomraning's time-independent asymptotic $P_N$ approximation.

\subsection{The Time-dependent Asymptotic $P_N$}
Finally, we derive the time-dependent analogy to Pomranin'g closure, using the time-dependant source free homogeneous medium Boltzmann equation. This method is a the generalization for Heizler's asymptotic $P_1$ approximation~\cite{heizler2010asymptotic}, to a general $N$. The derivation is shown here briefly, focusing on the new features (for further details see~\cite{harel2020time}). 

In order to derive the time-dependent asymptotic $P_N$ we start with Pomraning's closure, but with a slight modification in which we take the variable $\alpha_n$ outside the space derivative:
\begin{equation}\label{atd_nequation}
        \frac{1}{c}\pdv{I_N(x, t)}{t} + \sigma_t I_N(x,t) + \left(\frac{N}{2N+1}\right)\pdv{I_{N-1}(x, t)}{x} + \alpha_N\left(\frac{N + 1}{2N + 1}\right) \pdv{ I_{N-1}(x, t)}{x} = 0.
\end{equation}
This modification reproduces the asymptotic diffusion Fick's law, in the time-independent case for $N=1$~\cite{harel2020time}.
Additionally, $\alpha_n$ will now be derived by using the \textbf{time-dependent} source-free Boltzmann equation asymptotic distribution. Solving this equation is done by first applying Laplace transformation to Eq.~\ref{transport_Eq}:
\begin{equation}\label{s_boltzmann}
    \left(\frac{s}{c} + \sigma_t\right)\hI(x,\mu) + \mu \pdv{ \hI(x,\mu)}{x} = \frac{\sigma_t c \weff }{2}\hE(x).
\end{equation}
Defining the following $s$-dependent coefficients:
\begin{subequations}\label{Hnotation}
    \begin{equation}
        \hsigma_t = \sigma_t + \frac{s}{c} 
    \end{equation}
    \begin{equation}
        \hweff = \frac{\sigma_s}{\hsigma_t} = \frac{\weff}{1 + \frac{s}{c\sigma_t}}  
    \end{equation}
\end{subequations}
and plugging them back to Eq~\ref{s_boltzmann} yields:
\begin{equation}\label{boltzmann_I}
    \hsigma_t \hI(x,\mu) + \mu \pdv{ \hI(x,\mu)}{x} = \frac{\hsigma_t c \hweff }{2}\hE(x) 
\end{equation}
Eq.~\ref{boltzmann_I} can be solved spatially in exactly the same manner as Eq.~\ref{boltzmman_indep}, but, in the Laplace domain. Therefore, the asymptotic solution is exactly the same except for the modified $s$-dependent coefficients:
\begin{equation}\label{TD_asymptotc_eq}
\hI{}_{,a}(x,\mu)\approx\frac{\hweff}{2}A_0\frac{e^{\hkappa\hsigma_tx}}{1+\mu\hkappa}+\frac{\hweff}{2}B_0\frac{e^{-\hkappa\hsigma_tx}}{1-\mu\hkappa}
\end{equation}
With $s$-dependent closed transcendental equation for the asymptotic eigenvalues:
\begin{equation}
\frac{2}{\hweff}=\frac{1}{\hkappa}\ell n\left(\frac{1+\hkappa}{1-\hkappa}\right)
\label{hkappa}
\end{equation}
Plugging the time-dependent asymptotic solution - Eq.~\ref{TD_asymptotc_eq} into Eq.~\ref{alphapomraning0} yields:
\begin{equation}\label{alpha_s}
    \alpha_N\to\halpha{}_{,N} = \frac{\int_{-1}^{1} \hI{}_{,a}(x,\mu) P_{N+1}(\mu) d\mu}{\int_{-1}^{1} \hI{}_{,a}(x,\mu) P_{N-1}(\mu) d\mu}= \frac{\int_{-1}^{1} \left(\frac{1}{1 + \mu \hkappa}\right) P_{N+1}(\mu) d\mu}{\int_{-1}^{1} \left(\frac{1}{1 + \mu \hkappa}\right) P_{N-1}(\mu) d\mu}
\end{equation}

Next, applying the Laplace transform to Eq. \ref{atd_nequation} using the $s$-dependent coefficients $\hsigma_t$ and $\halpha{}_{,N}$ yields:
\begin{equation}\label{atd_nequation_laplace1}
    \hsigma_t\hI{}_{,N}(x) + \left(\frac{N + (N+1)\halpha{}_{,N}}{2N+1}\right)\pdv{\hI{}_{,N-1}(x)}{x} = 0
\end{equation}
This Fick's law-form equation, defines the $s$-dependent ``diffusion" coefficient:
\begin{equation}\label{asymptotic_pn_ficklaw}
        \hI{}_{,N}(x) = -\hDN(x) \pdv{\hI{}_{,N-1}(x)}{x} \quad\longrightarrow\quad \hDN(x) = \frac{N+(N+1)\halpha{}_{,N}}{(2N+1)\hat{\sigma}_t^s}
\end{equation}

As a generalization of Heizler's asymptotic $P_1$ approximation~\cite{heizler2010asymptotic}, an approximate closure (last $N$ equation), is derived by using two new coefficients $\A_N(\weff)$ and $\B_N(\weff)$: 
\begin{equation}\label{atd_nequation_laplace2}
    \frac{\A_N(\weff)}{c} \pdv{I_N(x, t)}{t} + \B_N(\weff) \sigma_t I_N(x,t) + \pdv{I_{N-1}(x,t)}{x} = 0.
\end{equation}
Further, applying the Laplace transform to the above approximated closure equation yields:
\begin{equation}\label{atd_nequation_laplace3}
    \hI{}_{,N}(x) = - \frac{c}{s\A_N(\weff) + c\sigma_t\B_N(\weff)} \cdot \pdv{\hI{}_{,N-1}(x)}{x}
\end{equation}
and by substituting the approximate Eq.~\ref{atd_nequation_laplace3} into the exact closure Eq.~\ref{atd_nequation_laplace1} yields:
\begin{align}
& \frac{(2N+1)(c\sigma_t+s)}{N+(N+1)\halpha{}_{,N}\left(\hkappa(\hweff)\right)}=\\
&\frac{(2N+1)(c\sigma_t+s)}{N+(N+1)\halpha{}_{,N}\left(\hat{D}_0(\hweff)\right)}\approx c\sigma_t\B_N(\weff) + \A_N(\weff)s + {\cal O}(s^2)\nonumber.
\label{AnBn}
\end{align}
The modified asymptotic coefficients, $\A_N(\weff)$ and $\B_N(\weff)$ can be solved with the integral notation for $\halpha{}_{,N}$ (Eq.~\ref{alpha_s}) under small-$s$ restrictions by using the Taylor expansion. $\halpha{}_{,N}$ is an explicit function of $\hkappa(\weff)$ or alternatively, $\hat{D}_0(\hweff)$, for each $\weff$. For different functional expression for $\hkappa(\weff)$ and $\hat{D}_0(\hweff)$ please see~\cite{case1953introduction} and  the appendix of~\cite{harel2020time}.

The details of the derivation for $N=1$, 2 and 3, and also the derivation of $\halpha{}_{,N}$ (which is practically finding the general solution for the integral in Eq.~\ref{alpha_s}) can be found in~\cite{harel2020time} (replacing $\weff$ in $c$) and we will not repeat it here. However we do mention, that for fully LTE problem, i.e. $\weff=1$, we reproduce the asymptotic $P_1$ results, $\A_1(1)=3/5$ and $\B_1(1)=3$~\cite{heizler2010asymptotic, heizler2012sp}. For asymptotic $P_2$ the method produces $\A_2(1)=4/7$ and $\B_2(1)=5/2$ and for asymptotic $P_3$ we yield $\A_3(1)=5/9$ and $\B_3(1) =7/3$. It is interesting to note that $\A_N(1)\to1/2$ and $\B_N(1)\to2$, when $N\to\infty$.

\begin{figure}[htbp!]
\includegraphics*[width=7.5cm]{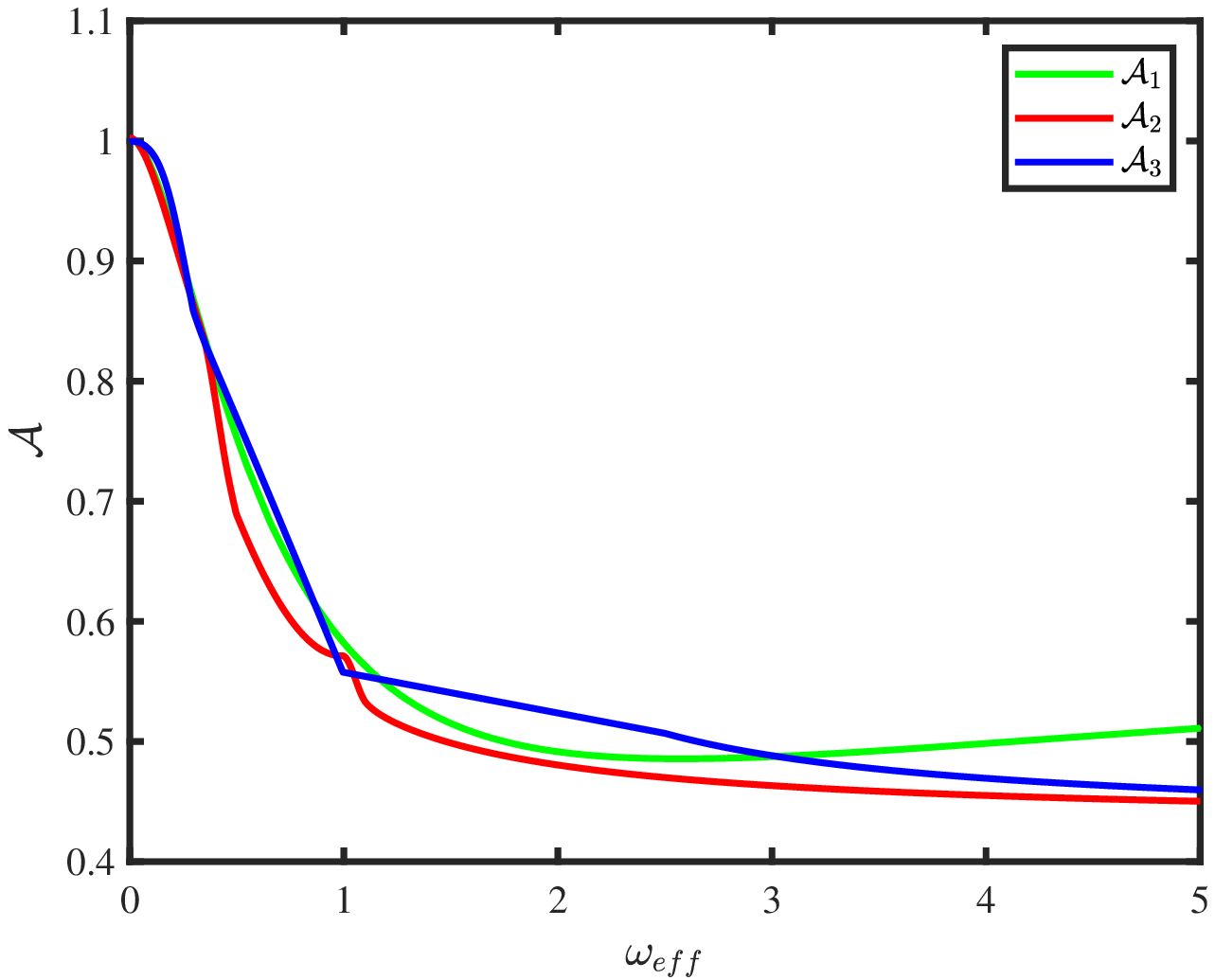}
\includegraphics*[width=7.5cm]{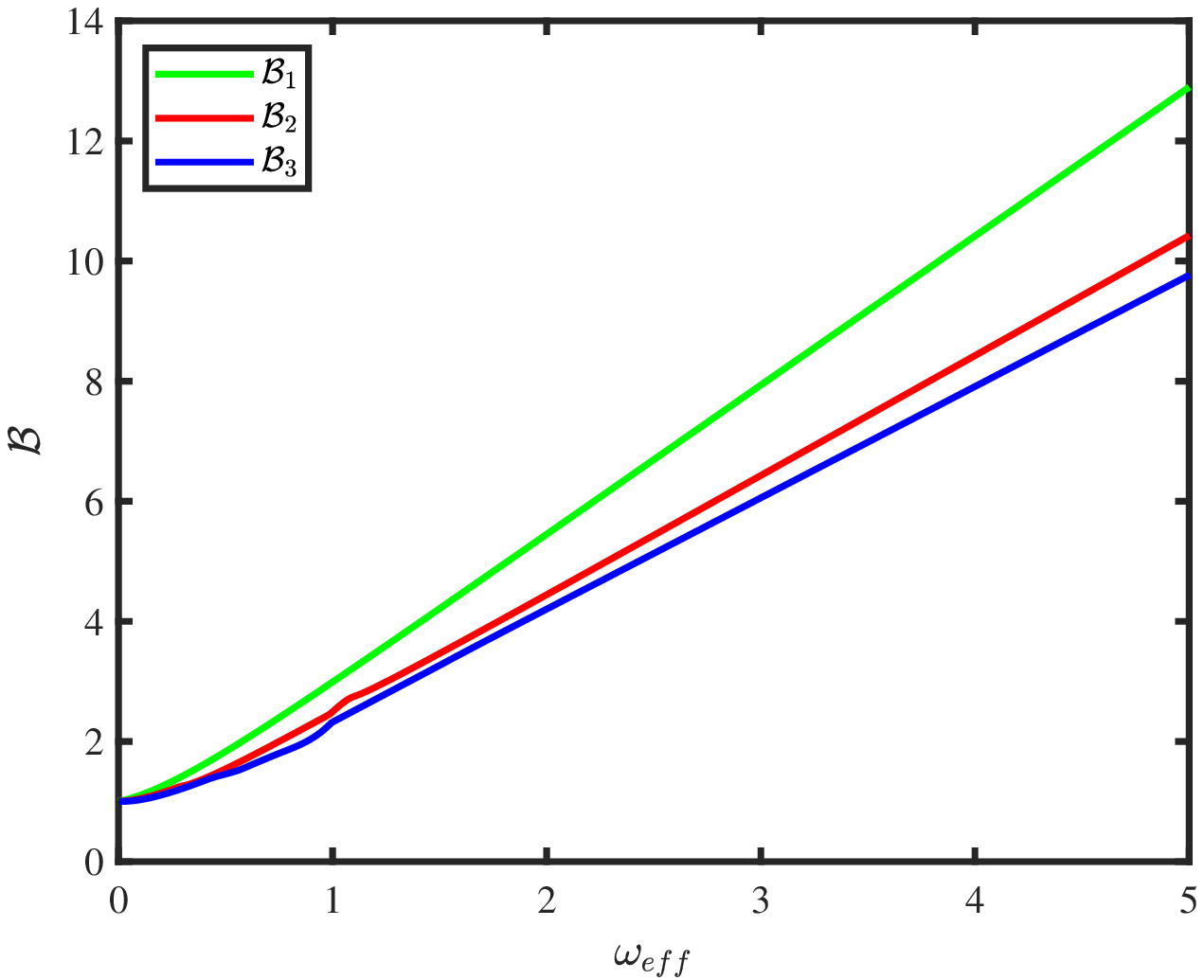}
\caption{The media-dependent coefficients $\A_N(\weff)$ and $\B_N(\weff)$ as function of the medium's albedo $\weff$ for the first 3 orders of the time-dependent asymptotic $P_N$ approximation ($N=1,2,3$).}
\label{fig:AB}
\end{figure}

In Fig.~\ref{fig:AB} we present the behavior of the coefficients $\A_N(\weff)$ and $\B_N(\weff)$ for different values of $\weff$ for the range of $0\leqslant\weff \leqslant 5$, here for the first time for values that are greater than 1. This regime is needed for strong sources, inner or outer, far from LTE regimes that are important in radiative transfer. It can be used also for neutronics in multiplying regions (in the presence of fission cross-sections). 

Fig.~\ref{fig:AB} suggests some important behaviors: First, $\B_N$ always equals to the inverse of Pomraning's time-independent asymptotic $P_N$ ``diffusion-like" coefficient, $\B_N=1/D_N(\weff)$. This forces the solution of the time-dependent asymptotic $P_N$ to tend to the solution of the time-independent asymptotic $P_N$ of Pomraning, when $t\to\infty$. Second, for fully-LTE medium ($\weff=1$), $\B_N(1)=3$ equals to the classic $P_N$ approximation coefficient for general $N$ but $\A_N(1)$ is different (as in Heizler's asymptotic $P_1$ has shown~\cite{heizler2012sp}). Third, for pure absorbing case (in the limit of cold non-LTE regions $\weff=0$), $\A_N(0)=\B_N(0)=1$ for any general $N$, yielding the exact particle velocity. We note that for $\weff>1$, the value of $\A_N$ is stabilized at $\A_N\approx 1/2$ for a wide range of $\weff$.

It should be noted that Pomraning proposed an approximate suggestion in~\cite{pomraning2005equations} (Eq. 3.147), which we call the $P_{1/\B_N}$ approximation. In this proposal, the {\em time-independent} closure is taken, therefore, determining $\B_N$ to be exactly as in the time-dependent asymptotic $P_N$ approximation:
\begin{equation}\label{pom_B}
\frac{1}{\B_N}\equiv D_N = \frac{N+(N+1)\alpha_N}{(2N+1)}
\end{equation}
when $\alpha_N$ is determined by Eq.~\ref{alphapomraning} and by setting $\A_N=1$. 
For the case of {\em{purely steady-state case}} $\weff=1$, this $P_{1/\B_N}$ approximation has even simpler form, and it is the $N$'s expansion of the $P_{1/3}$ approximation, which was offered by Olson~\cite{olson2012alternate}, $\A_N=1$ and $\B_N=(2N+1)/N$ (which is the coefficient of the classic $P_N$ approximation when setting $\alpha_N=0$ in Eq.~\ref{pom_B}). The $P_{1/\B_N}$ approximation yields the exact particle velocity by definition due to the $\A_N=1$ choice. When approaching to the steady state, and as $N\to\infty$,  the asymptotic $P_N$ and the $P_{1/\B_N}$ approximations yields closer results. This means that the time-dependent asymptotic $P_N$ provides the rigorous mathematical justification (due to asymptotic analysis derivation) for the $P_{1/\B_N}$ approximation.

The results of Fig.~\ref{fig:AB} will serve as an input for any benchmark or simulation. In each zone (which in radiative transfer is any  numerical cell) $\weff(\vec{r},t)$ is calculated (via Eq.~\ref{omega}) for each time-step, due to the temperature dependency of the opacity and the difference between the material and radiation energies. Then we set $\A_N(\vec{r},t)$ and $\B_N(\vec{r},t)$ for each numerical cell in a given time-step by solving the linear set of $P_N$ equations, or simplified $P_N$ ($SP_N$) in multi-dimensions~\cite{spn}.

\section{Su-Olson Benchmark}\label{test_benchmark}

In order to test our time-dependent asymptotic $P_N$ approximation in its radiative form, we used the well-known Su-Olson benchmark ~\cite{su1997analytical}. The Su-Olson benchmark is a non-equilibrium gray one-dimensional radiative transfer source problem, with a constant opacity in an infinite medium.
\begin{figure}[htbp!]
\includegraphics*[width=8cm]{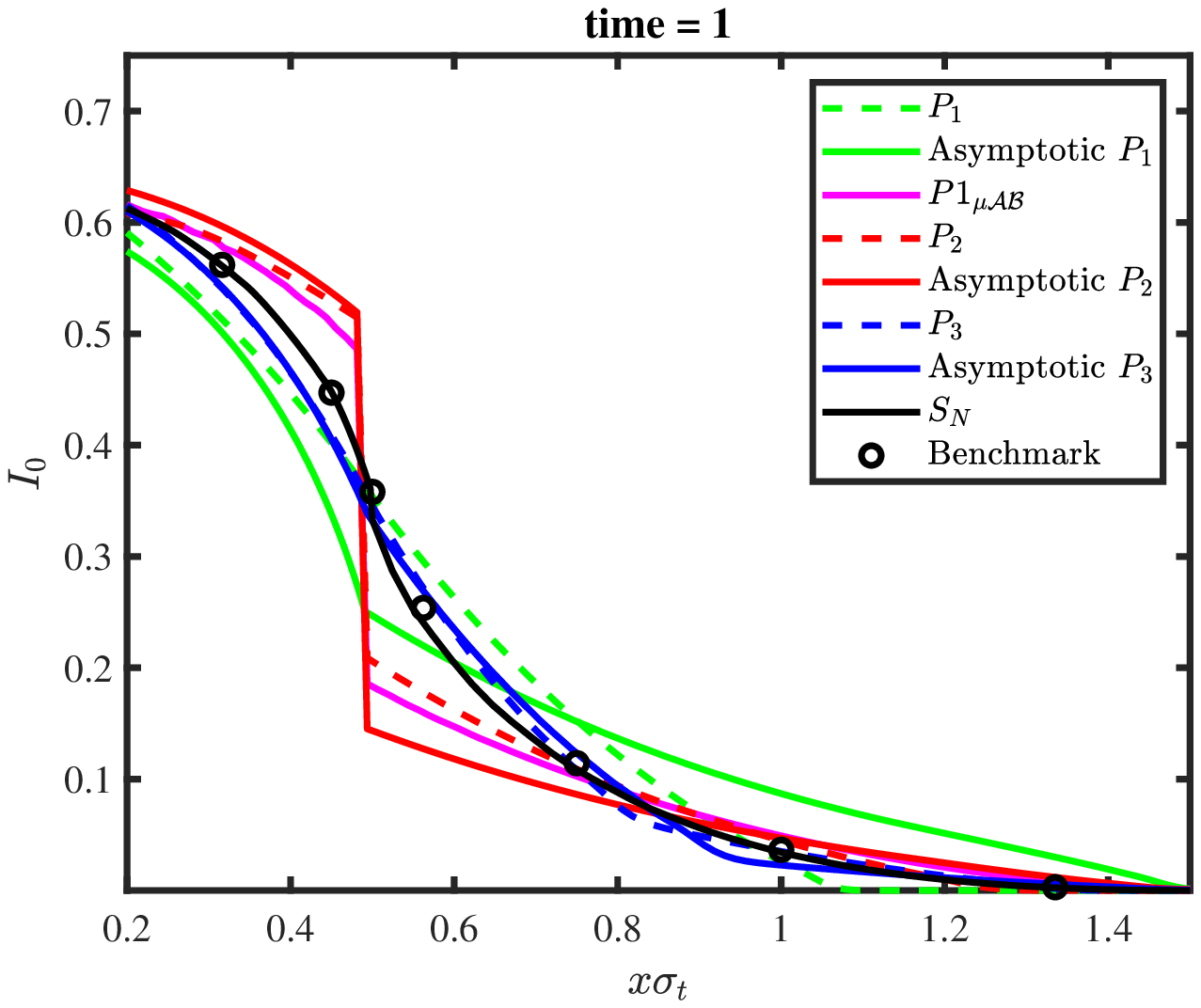}
\includegraphics*[width=8cm]{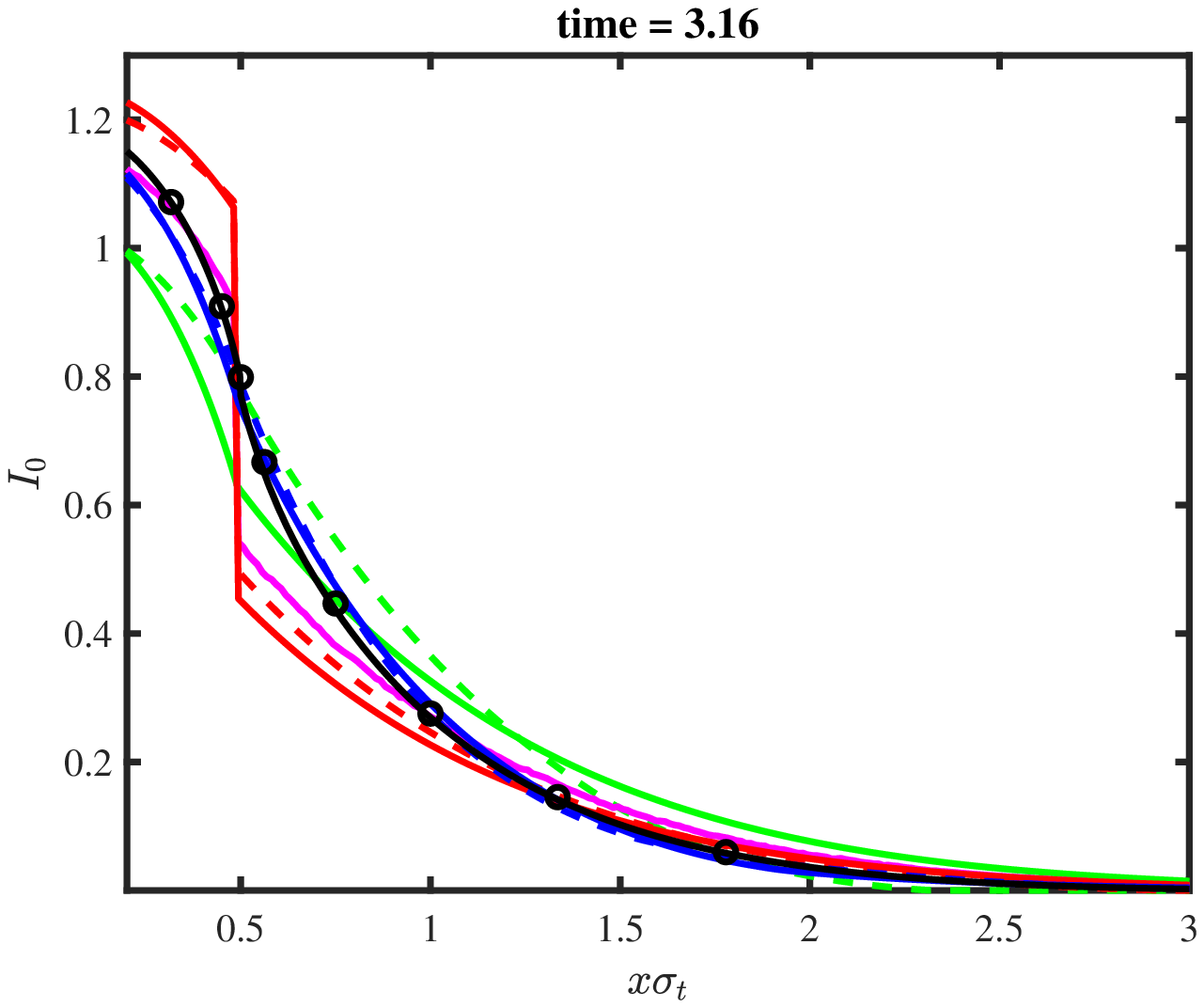}
\includegraphics*[width=8cm]{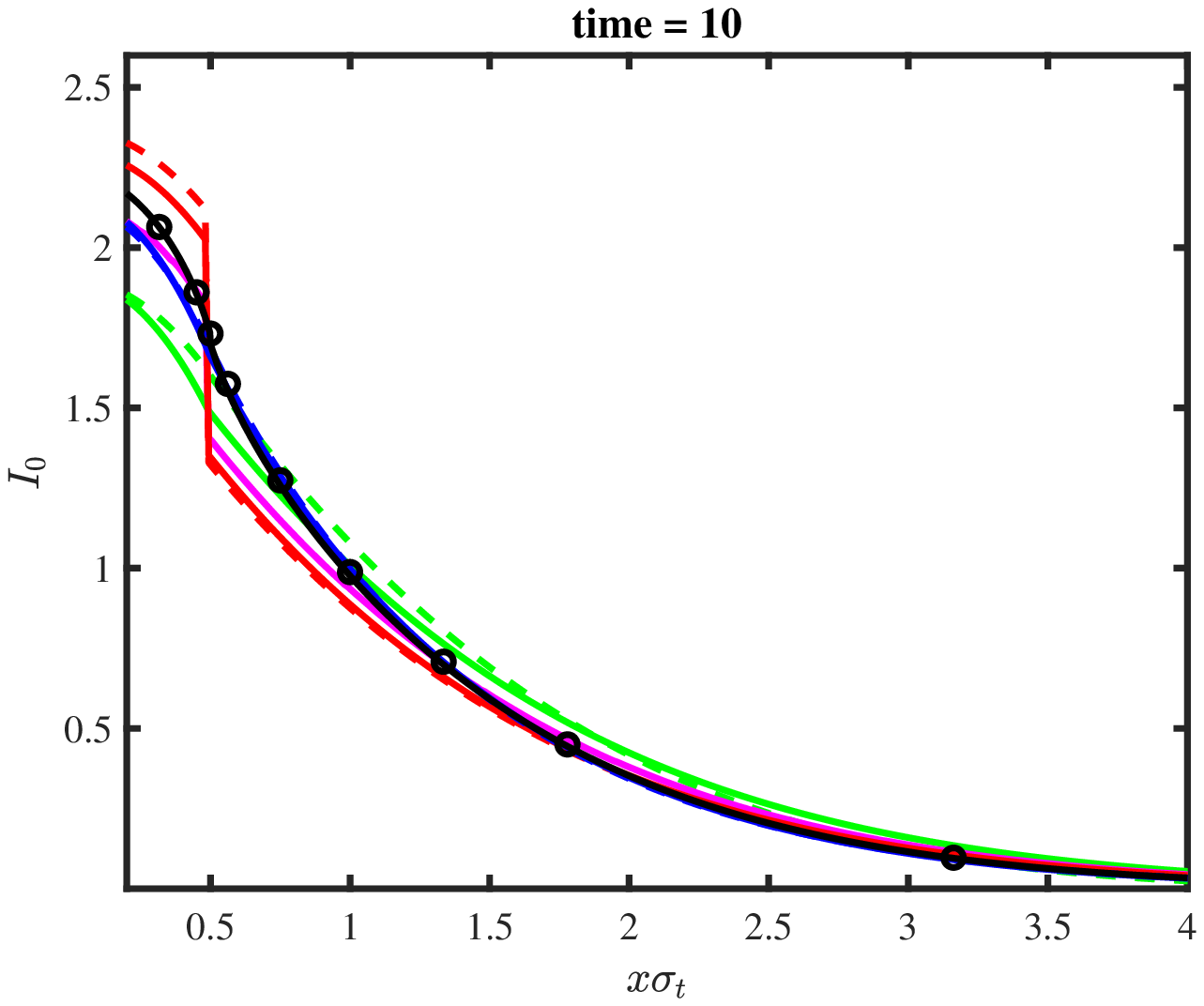}
\caption{The radiation energy for the case of $c_s=0$ in linear scale. The different classic $P_N$ approximations are presented in dashed curves, while the asymptotic $P_N$ are presented by the solid curves. The exact benchmark results are shown in the black circles, while the $S_N$ results using $N=64$ are presented by the black curves. We have added the discontinuous asymptotic $P_1$ results from~\cite{cohen2018discontinuous} (the magenta curves).}
\label{fig:ca1_linear}
\end{figure}
\begin{figure}[htbp!]
\includegraphics*[width=8cm]{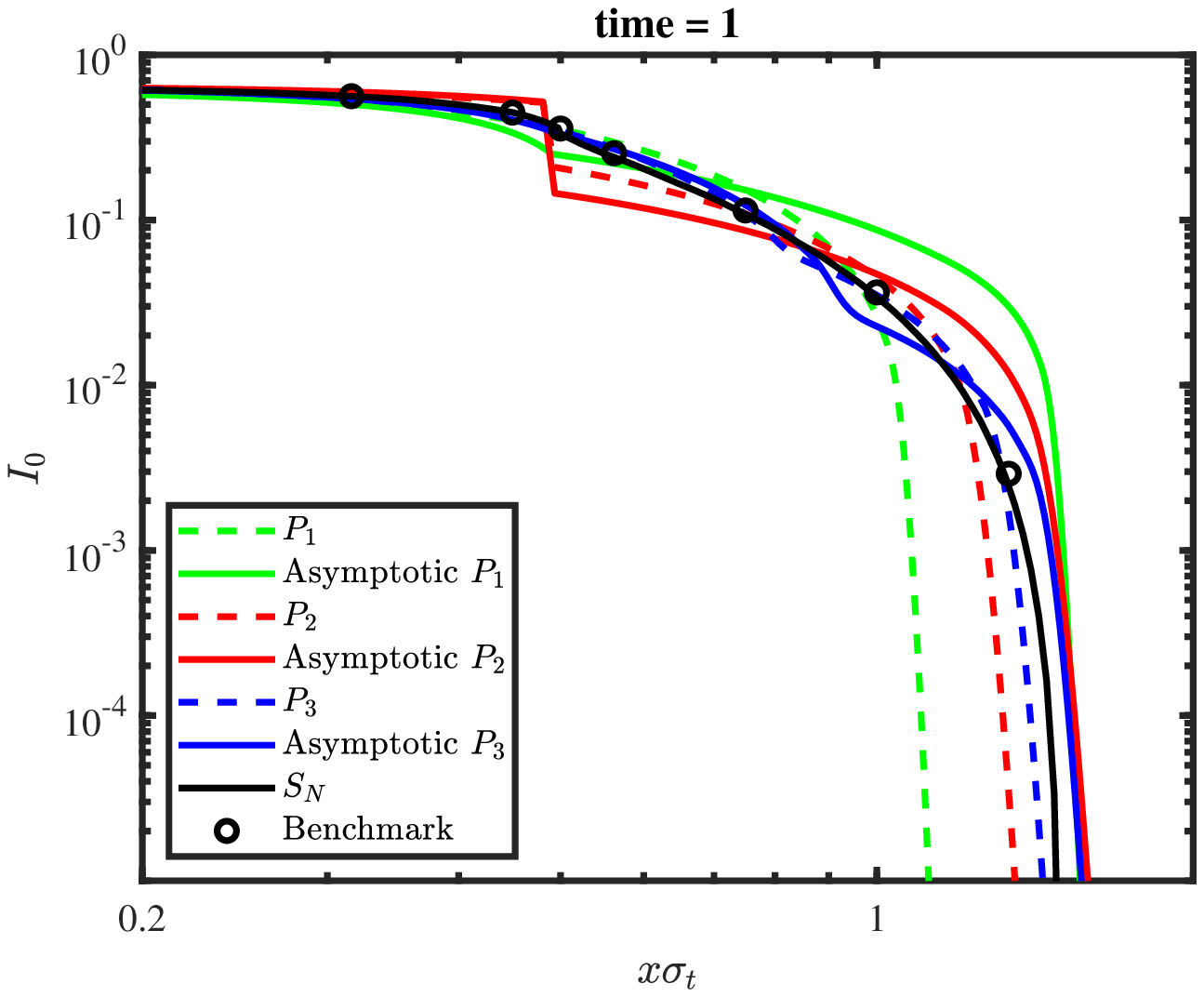}
\includegraphics*[width=8cm]{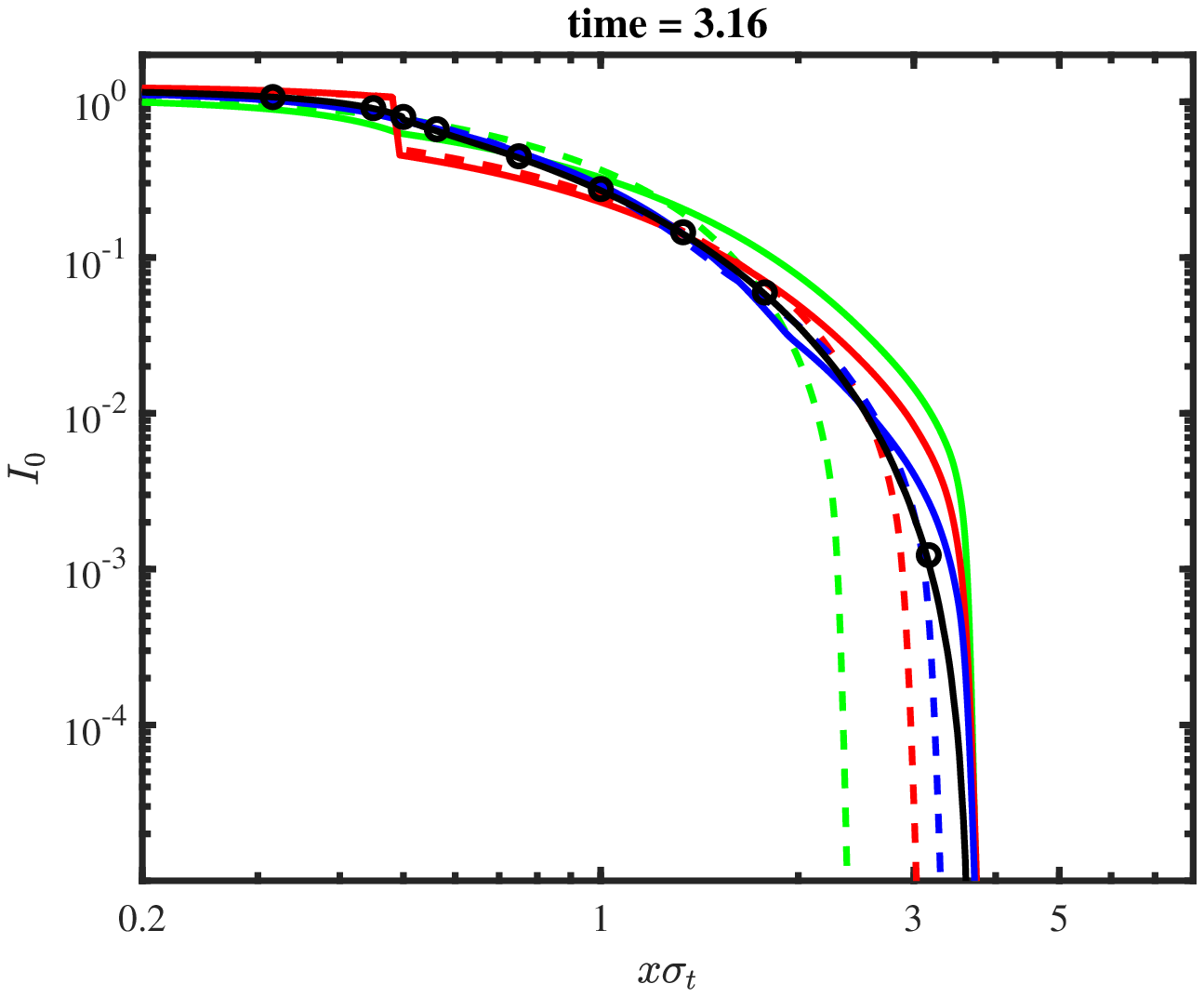}
\includegraphics*[width=8cm]{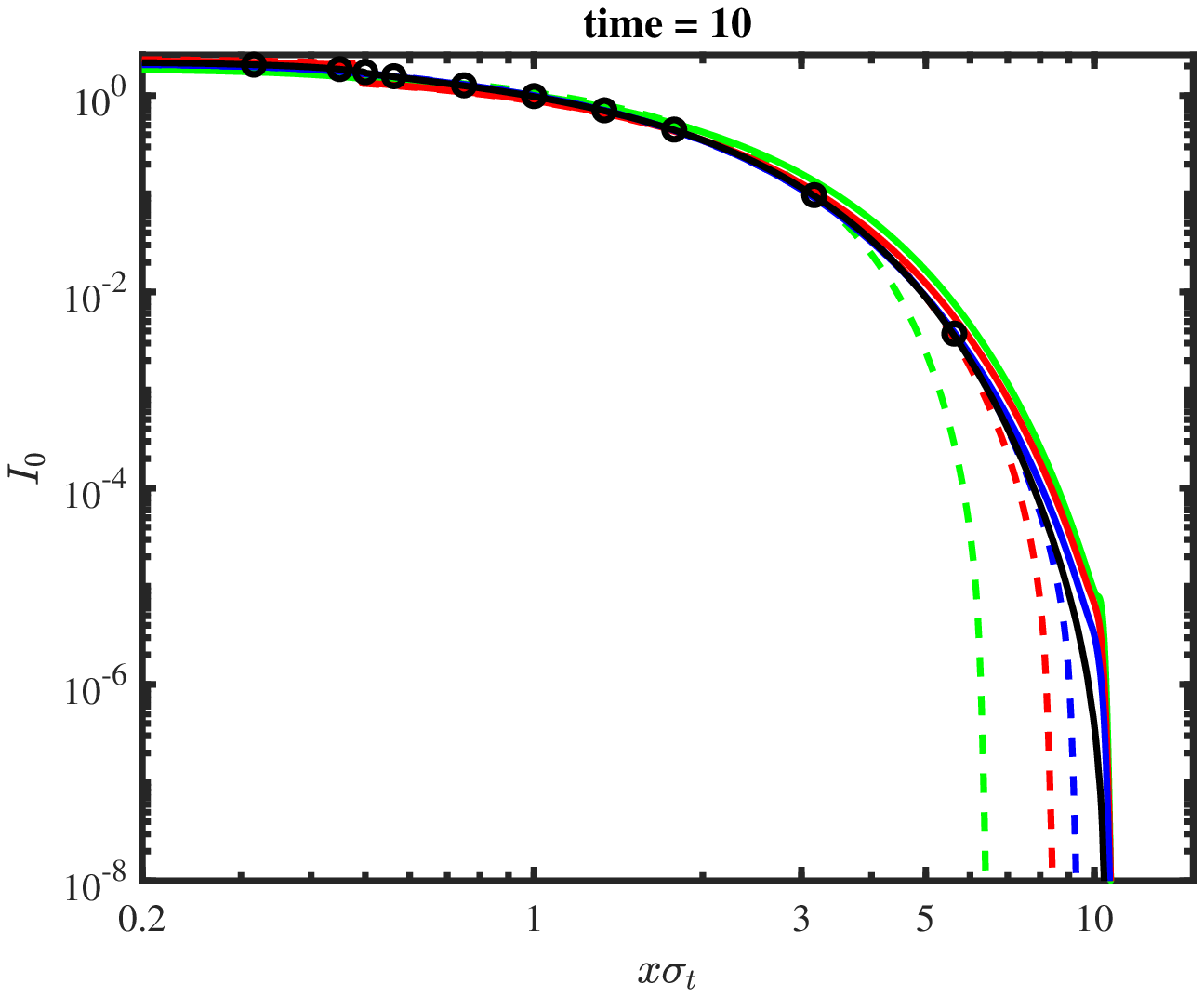}
\caption{The radiation energy for $c_s=0$ in log scale.}
\label{fig:ca1_log}
\end{figure}

It is common to present the physical properties as dimensionless as done by Cohen et al.~\cite{cohen2018discontinuous}, i.e. in dimensionless space units $z\equiv\sigma_{t} x$ and dimensionless time $\uptau\equiv\epsilon c \sigma_{t} t$. The material heat capacity is defined as: $C_v=\alpha T^3$ and $\epsilon=4a/\alpha$. It is also convenient to define the ratio of the scattering cross section to the total cross section $c_s=\sigma_s/ \sigma_{t}$, since we use dimensionless position variable.
The radiation source for this benchmark is isotropic and constant for $z \leqslant 0.5$ and $\uptau \leqslant 10$:
\begin{equation}
S(z,\uptau)=\begin{cases}
      1, & \text{if}\ \uptau\leq 10, \quad z\leq 0.5 \\
      0, & \text{otherwise}
    \end{cases}
    \label{source}
\end{equation}
We compared both the time-dependent asymptotic $P_N$ approximation to the classical $P_N$ approximation and included the benchmark analytic solutions and $S_N$ solution with $N=64$.
The numerical simulations for both the classical and the asymptotic $P_N$ approximation, were executed in a fully implicit scheme for time, with a finite difference scheme for space in {\em{Matlab}} (we have used the \textit{sparse} option for accelerating the band-matrix inversion). We used a constant spatial resolution of $\Delta z\approx 1.25\cdot 10^{-3}$ while $\Delta \uptau$ is defined dynamically such that the energy intensity will not change in each cell (between time steps) more than 0.05\%.

The first case that we used to test our approximation is the $c_s=0$, i.e. purely absorbing media in the absence of scattering. It is important to note that as opposed to neutronics, purely absorbing media does {\em{not}} mean $\weff=0$. In radiative transfer $\weff$ is determined by the ratio of the material energy to the radiation energy due to the black-body radiation source (and the external source as well). So in reaching LTE when $E\approx aT_m^4$ for sufficiently long times, $\weff\to1$ even in the purely absorbing case, due to the absorption-emission process.
\begin{figure}[htbp!]
\includegraphics*[width=8cm]{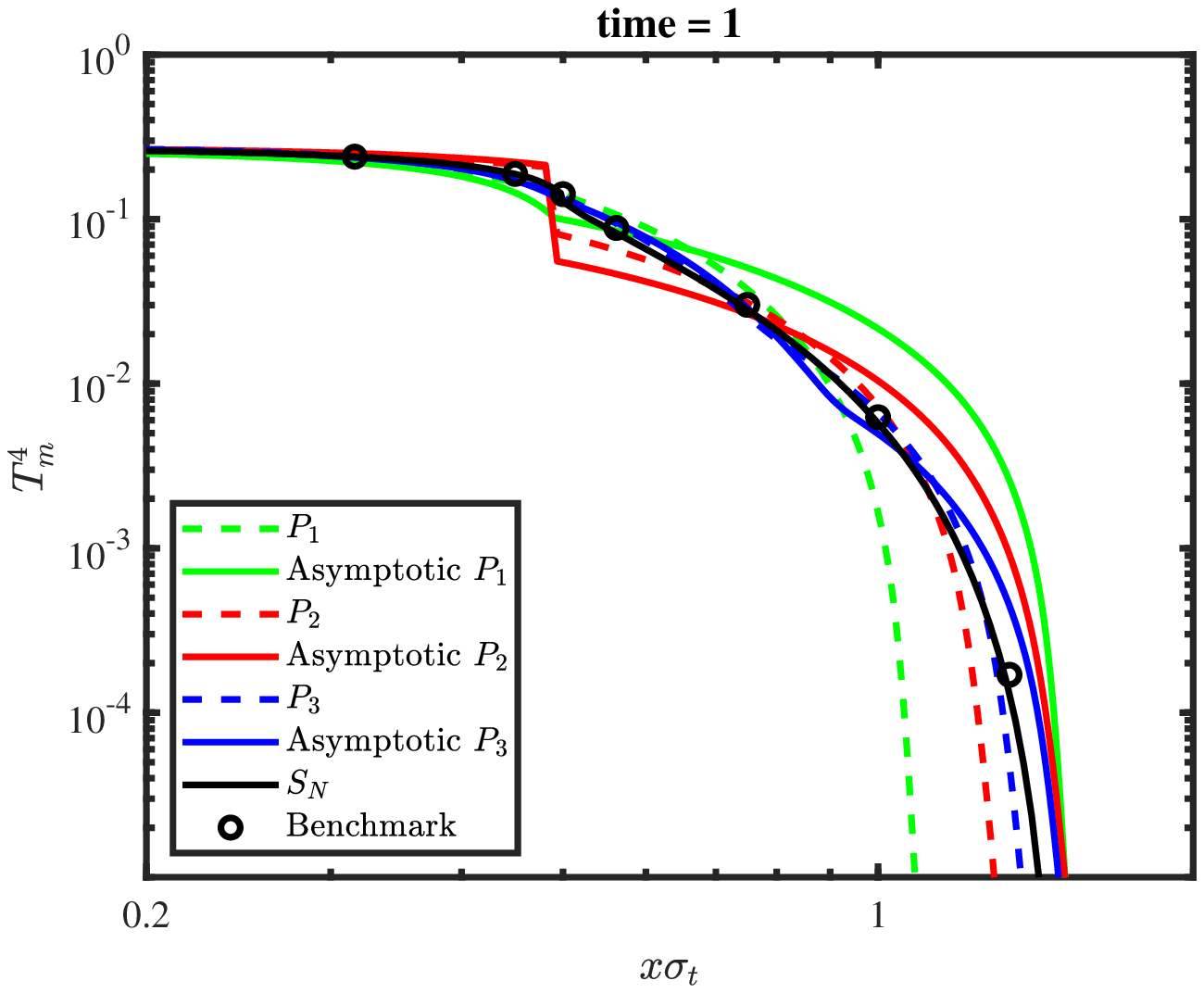}
\includegraphics*[width=8cm]{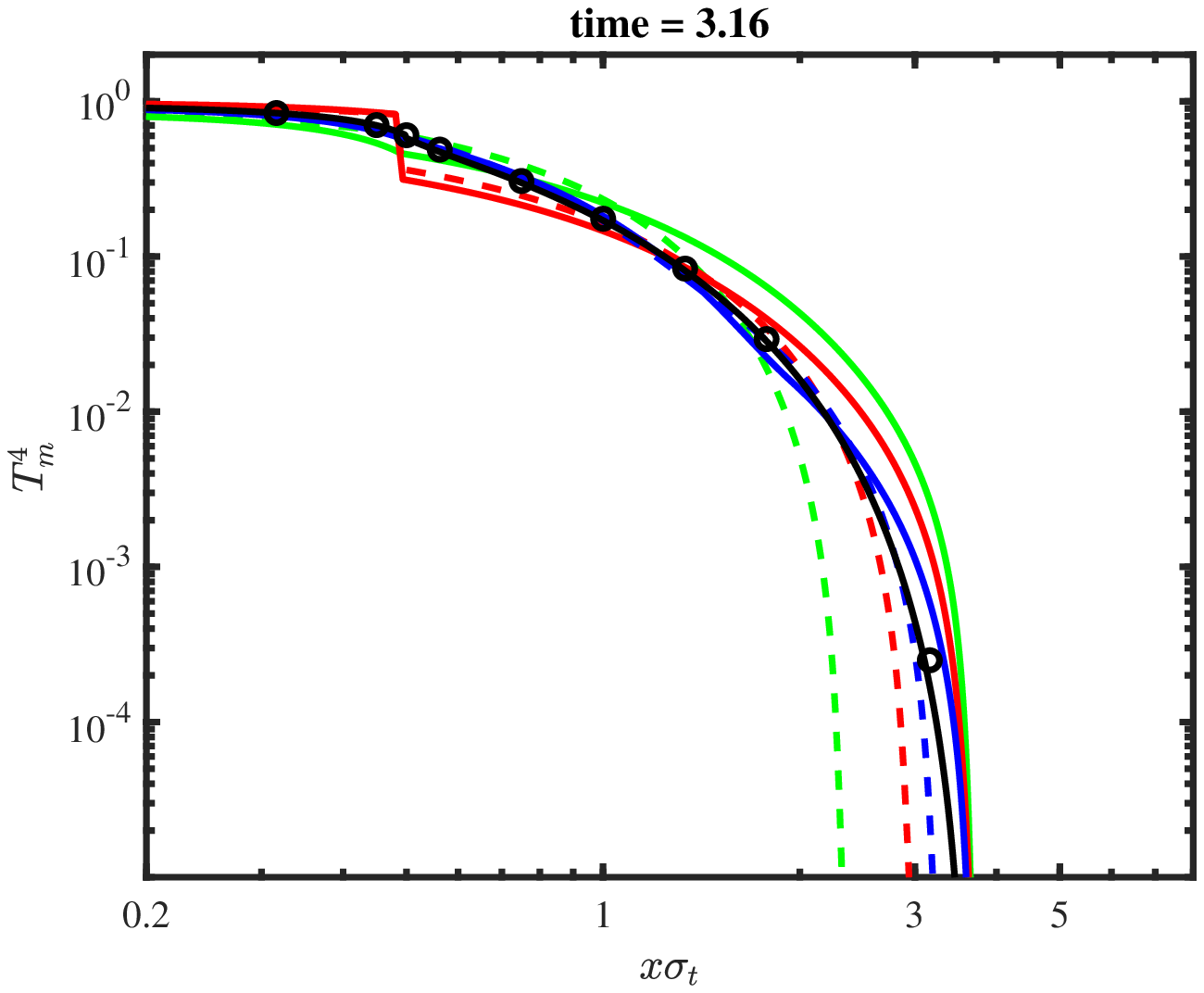}
\includegraphics*[width=8cm]{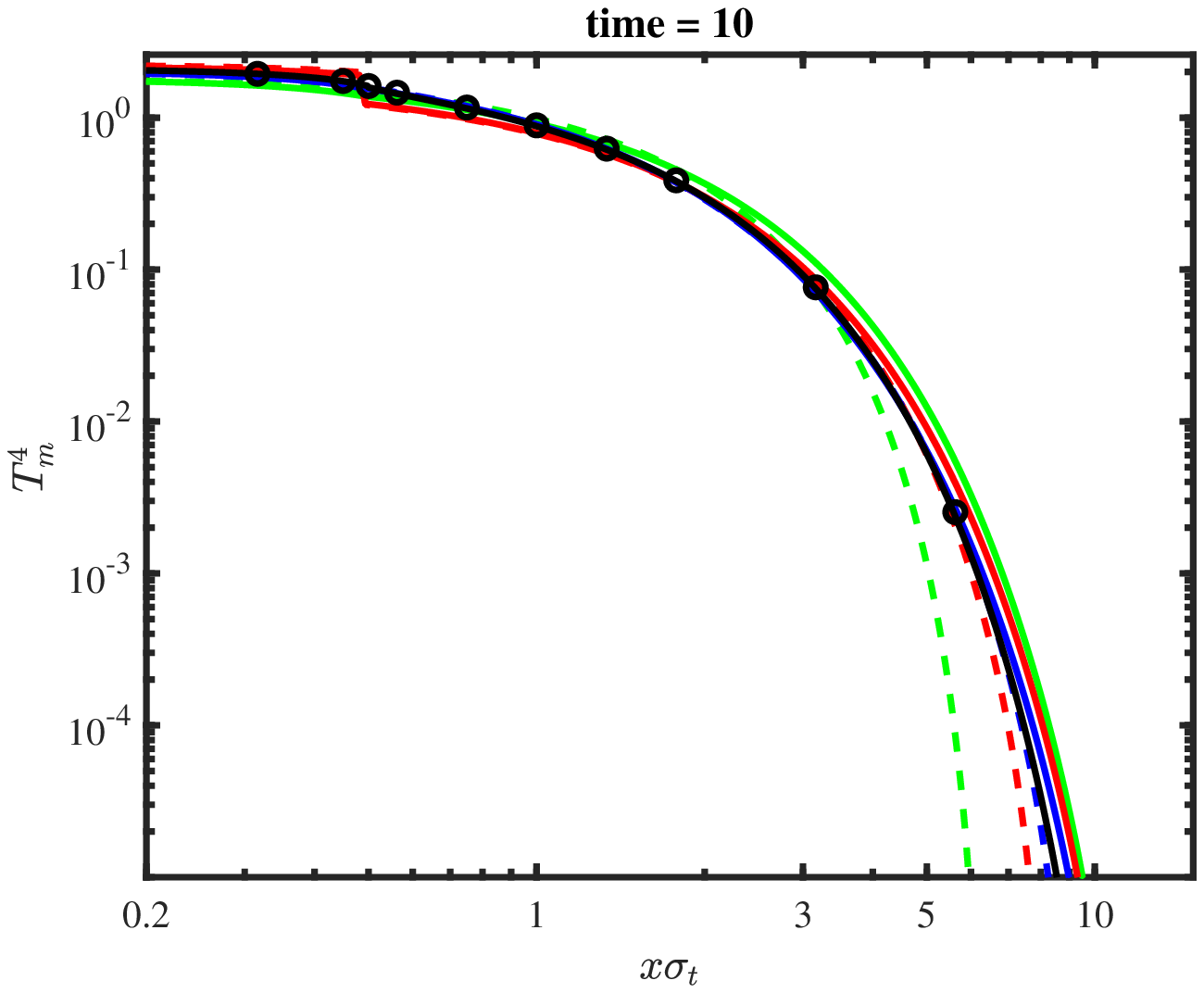}
\caption{The material energy for $c_s=0$ in log scale.}
\label{fig:mat_ca1_log}
\end{figure}

In the following figures, the radiation energy as a function of the dimensionless space, is presented, where Fig.~\ref{fig:ca1_linear} is in a linear scale and Fig.~\ref{fig:ca1_log} is in a logarithmic scale. The linear scale focuses in the radiation in the bulk regions, while using the logarithmic scale we can test the tails of the distribution. The results are shown for $t=1,3.16$ and 10 dimensionless times. The different classic $P_N$ approximations are presented in dashed curves, while the asymptotic $P_N$ are presented by the solid curves. The exact benchmark results are shown in the black circles, while the $S_N$ results using $N=64$ are presented by the black curves.

As one can see, the benchmark exact solution and the $S_{64}$ fits perfectly, both in the bulk and in the tails. In the bulk regions, near the source (Fig~\ref{fig:ca1_linear}), the difference between classic and asymptotic $P_N$ is small, for any given $N$.
The major difference is near the tails of the distribution, as can be seen in the logarithmic scale (Fig~\ref{fig:ca1_log}). Each of the classical $P_N$ solutions yields a heat front that is too slow while the asymptotic time-dependent solution results in a heat front that are slightly faster. However, it is clear that the results produced by the asymptotic time-dependent approximation are better than the classical approximation, yielding almost the correct particle velocity. 

Interesting feature is that the $P_2$ (both asymptotic and classic) have a discontinuous {\em{jump}} at the source surface, in $z=0.5$. This result, of the discontinuity of $P_2$ is well-known, due to the perpendicular direction of even $P_N$'s~\cite{RulkoLarsen1993}. It is quite similar to the jump of the discontinuous asymptotic $P_1$ results (the magenta curves)~\cite{cohen2018discontinuous}, however, the discontinuous asymptotic $P_1$ yields better results than both classic or asymptotic $P_2$, especially in long times. We note that such a discontinuity in the energy density is non-physical, since the exact solution of the RTE is of course, continuous. However, the physical justification of using discontinuities on the boundary between two different media rises, as mentioned above, from the physics of neutron transport. Discontinuous asymptotic diffusion theory ($P_1$ approximation) yields the correct critical radii in reactor-reflector systems, while the classic $P_N$ converges slowly compared to the exact solution~\cite{doyas}. This means that the non-physical existence of discontinuity is negligible compared to the benefits of correctly describing the particle's distribution  far from the boundary.

In Fig.~\ref{fig:mat_ca1_log} the complimentary results of the material energy is presented. The results of the material energy is similar qualitatively to the radiation energy results, of course differs quantitatively. As with the radiation energy, our new approximation yields better results than the classical one.

\begin{figure}[htbp!]
\includegraphics*[width=8cm]{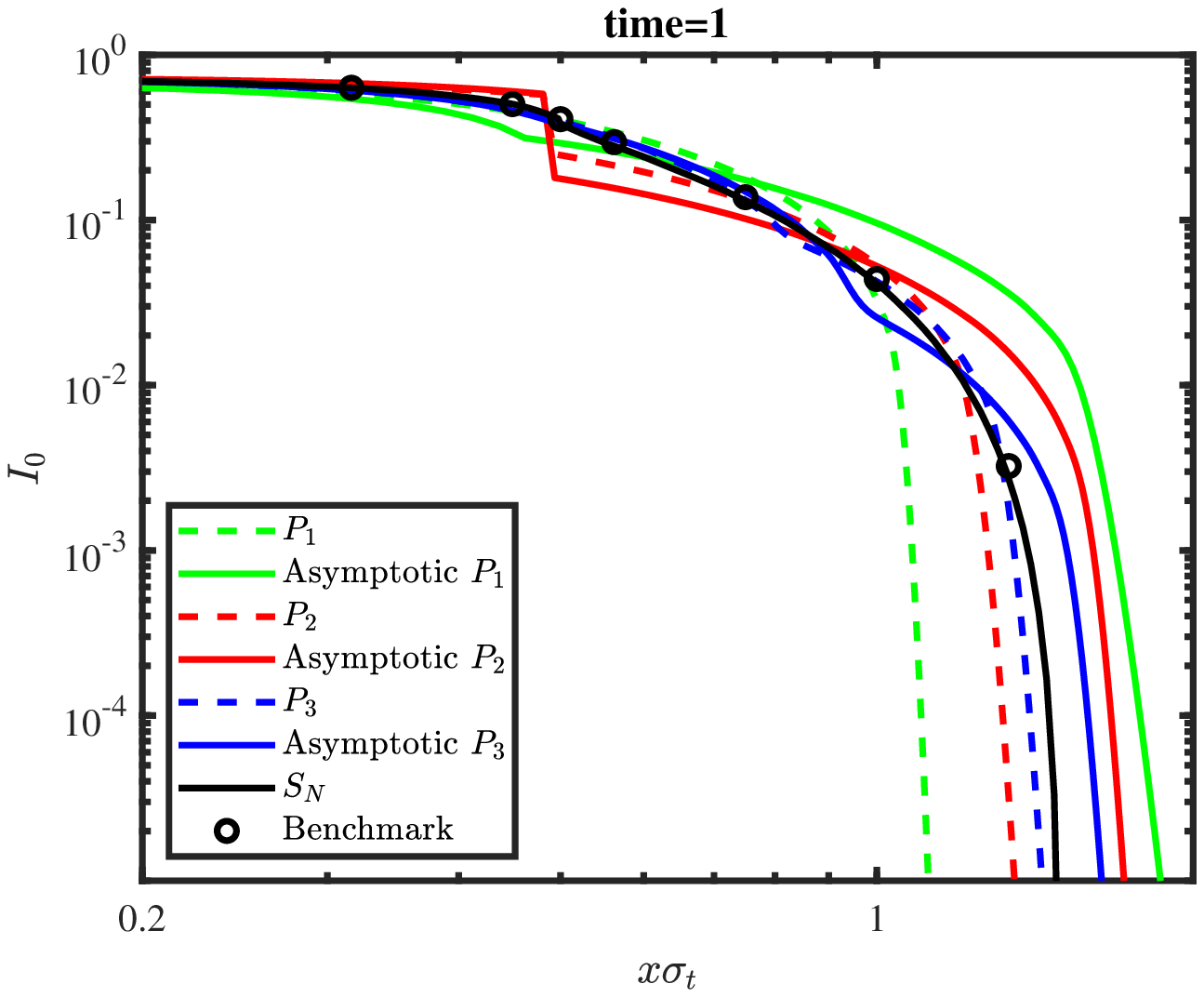}
\includegraphics*[width=8cm]{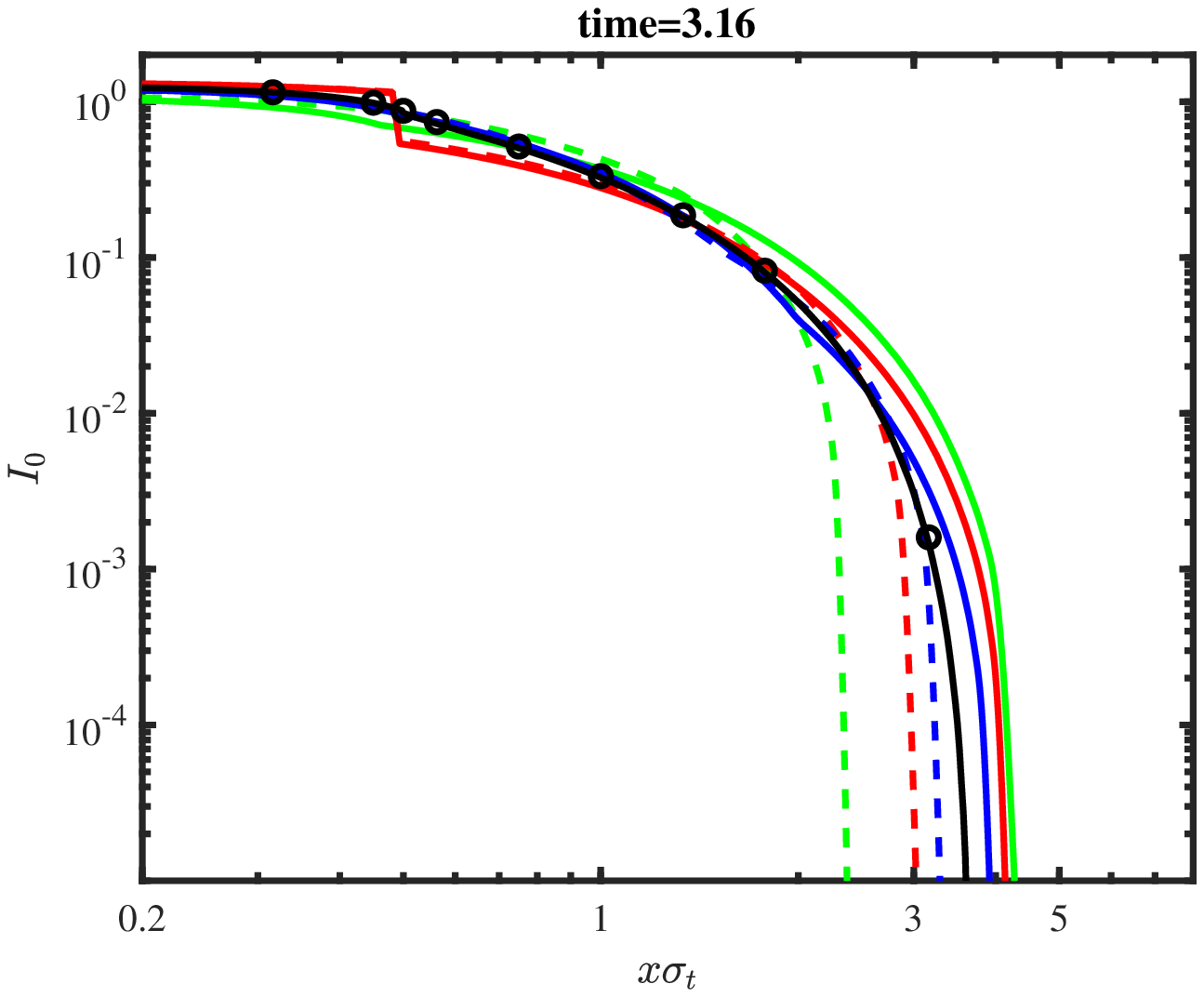}
\includegraphics*[width=8cm]{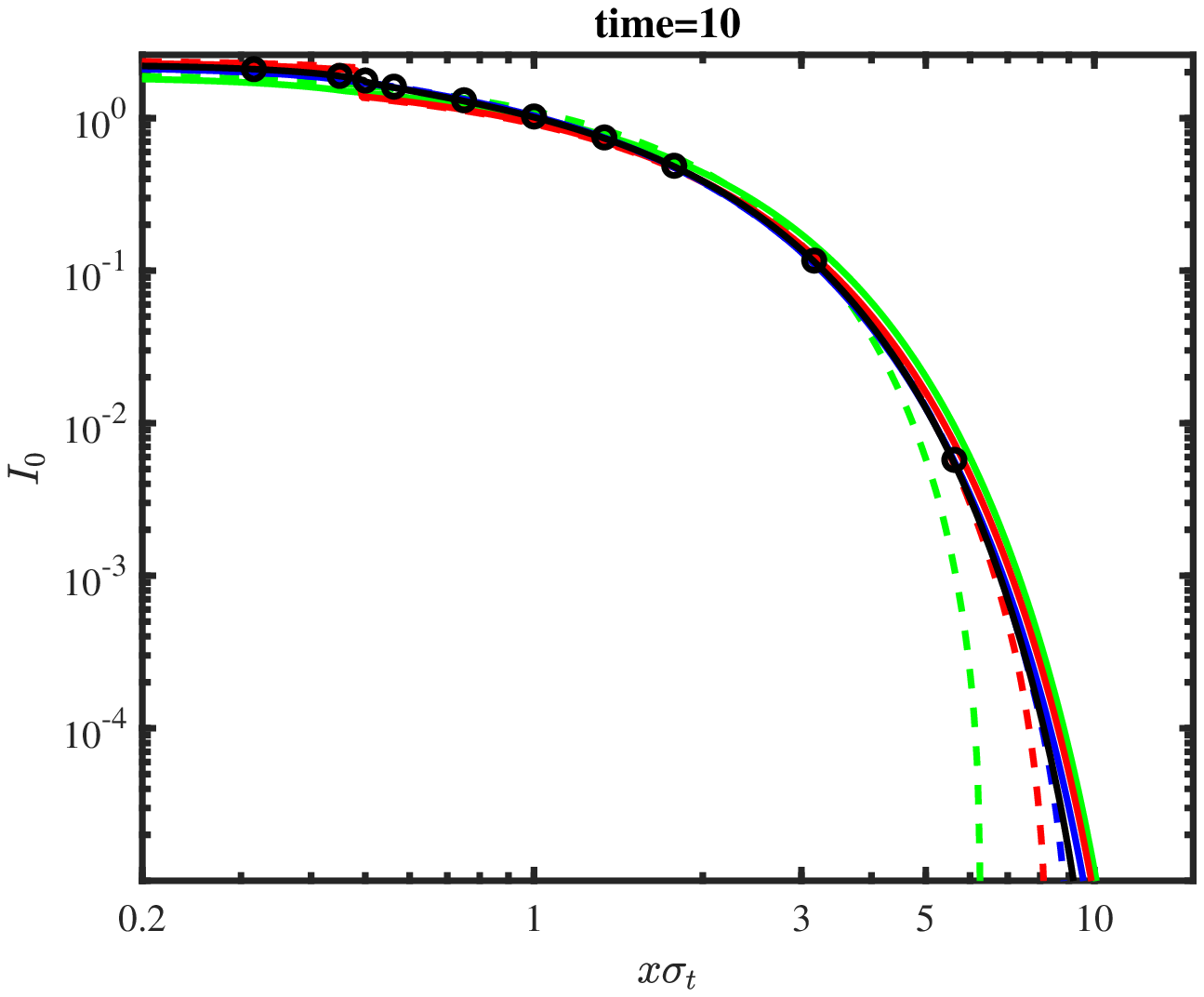}
\caption{The radiation energy for $c_s=0.5$ in log scale.}
\label{fig:ca05_log}
\end{figure}
We also tested our new approximation for the scattering-included term of $c_s=0.5$. The following Fig~\ref{fig:ca05_log} presents the results. Similar conclusions regarding $c_s=0$ case, is valid for this one, however, the asymptotic $P_N$ results in this case are a little bit worse than in the pure absorbing case (mostly in early times), near the front of the heat waves.

\section{Discussion}
In this study, we presented the time-dependent asymptotic $P_N$ approximation for radiative transfer. The asymptotic $P_N$ approximation is based on the $P_N$ approximation and Pomraning's time-independent asymptotic closure method~\cite{pomraning1964generalized}, and was presented in previous work for general transport problems~\cite{harel2020time}. In this study we have focused on the modifications that had to be derived for radiative transfer problems, and the non-LTE radiative transfer benchmark problem, for which we examined this approximation. 

The asymptotic $P_N$ approximation was derived in the same manner as all $P_N$ approximations are derived - by expanding the specific intensity into Legendre polynomials. Afterwards we introduced a new closure equation that is similar to Pomraning's closure.
As a generalization of the asymptotic $P_1$, the asymptotic $P_N$ approximation uses two media-dependent coefficients $\A_N(\weff)$ and $\B_N(\weff)$ {\em{in the closure equation}} that rests on the asymptotic distribution of the exact transport solution for infinite homogeneous medium, both in space (using Case et al., methodology~\cite{case1953introduction,CaseZweifel1967}) and time (using the Laplace domain in small-$s$). The coefficients $\A_N(\weff)$ and $\B_N(\weff)$ were derived for values of $\weff \leqslant 5$, thus enabling radiative transfer source problems in which $\weff$ may be in some cases, larger than unity (which may be also efficient for multiplying media in neutronics).

We have tested the new time-dependent asymptotic $P_N$ approximation for the well known Su-Olson benchmark in two cases - pure absorbing case ($c_s=0$) and scattering-included case $c_s=0.5$. The results show that our approximation share similar results to the classic $P_N$ in the bulk (near source) regions, and yields better results than the classical $P_N$ for any given $N$ near the tails of the distributions (the heat fronts regions). One important conclusion emerge from the results: the classic and the asymptotic $P_N$ approximations can be used as ``bounds" for the exact transport distribution, at least in the tails regions. The classic $P_N$ converges to the exact solution from the {\em{slower}} side, while the asymptotic $P_N$ converges from the {\em{faster}} side (though, converges faster in $N$ the classic $P_N$).

An interesting property, both the time-dependent $P_2$ (both classic and asymptotic) and the discontinuous asymptotic $P_1$ includes a jump at $z=0.5$. Although such discontinuity itself in the radiation energy density is of course non-physical, there are  benefits for describing correctly the  distribution of the particles far from the boundary.

In the future, we plan to extend our work to multi-dimensional problems, via the simplified $P_N$ ($SP_N$) approximation~\cite{spn}. Specifically, we are interested to explore the supersonic Marshak wave experiments, when a heat wave propagates into a low-Z foam, coated with a high-Z envelope (e.g. Au)~\cite{prr}. Since the coefficients $\A_N(\weff)$ and $\B_N(\weff)$ depend on the properties of numerical cells (i.e.~material), via $\weff(t,\vec{r})$ (which is function of time and space) and free of gradient (spatial derivatives) or time derivative, the extension to multi-dimensions is straightforward by using the existent time-dependent $SP_N$ code, which for low-order $N$, should yield a sufficiently good accuracy (same should work for other one-dimensional geometries). The only modification that is needed is setting the coefficients of the last (closure) equation to $\A_N(\weff)$ and $\B_N(\weff)$, that were derived and presented in this study (for example in Fig.~\ref{fig:AB}), instead of the classic $P_N$ coefficients.

\begin{acknowledgments}
We acknowledge the support of the PAZY Foundation under Grant \textnumero~61139927.
\end{acknowledgments}
\setlength{\baselineskip}{12pt}
\bibliographystyle{unsrtnat} 
\bibliography{bibliography} 

\end{document}